\documentclass[options]{JHEP3}
\title{That's a wrap!}
\author{Tasneem Zehra Husain\\ Department of Physics,\\
Stockholm University,\\
PO Box 6730,\\
S 11385 Stockholm,\\
Sweden.\\
Email: \email{tasneem@physto.se}}
\abstract{
Calibration technology provides us with a fast and elegant way to find the 
supergravity solutions for BPS wrapped M-branes. Its true potential 
had however remained untapped due to the absence of a classification of 
calibrations in spacetimes with non-trivial flux. The applications of this method
were thus limited in practise to M-branes wrapping Kahler calibrated cycles. In this 
paper, we catagorize a type of generalised calibrations which exist in supergravity 
backgrounds and contain Kahler calibrations as a sub-class. This broadens the arena of 
brane configurations whose supergravity solutions are accessible through the 
calibration 'short-cut' method.} 


\preprint{USITP-02-04\\ hep-th/0302071}

\newcommand{\C}{{\mathbb C}}

\newcommand{\comment}[1]{}

\def\bbbz{{\sf Z\!\!\!Z}}
\def\sl2z{SL(2,\bbbz)}
\newcommand{\be}{\begin{equation}}
\newcommand{\ee}{\end{equation}}
\newcommand{\bea}{\begin{eqnarray}}
\newcommand{\eea}{\end{eqnarray}}

\def\bbbz{{\sf Z\!\!\!Z}}
\def\sl2z{SL(2,\bbbz)}

\def\z0{{\bf z_0}}

\newcommand{\bit}{\begin{itemize}}
\newcommand{\eit}{\end{itemize}}
\newcommand{\no}{\noindent}
\begin{document}
\setcounter{page}{1}
\pagestyle{plain}

\section{Introduction}

Flat M-branes are solitons which preserve half the spacetime supersymmetry 
and are charged under the three-form of 11-dimensional supergravity.
The BPS spectrum of M-Theory includes not just flat branes but also states 
corresponding to intersecting and wrapped brane configurations, the supergravity
solutions of which have been widely discussed in recent years. (For reviews of the
subject, see \cite{Lecnotes}). In spacetimes where the supergravity three-form 
is constant, supersymmetry preservation demands that compactification manifolds 
have special holonomy and the calibrations which exist on such manifolds are known. 
In more 'realistic' situations however, where the supergravity field strength 
is non-zero, 
we do not yet have an exhaustive list of possible calibrations, though relevant work 
has been done recently in \cite{Gst} and \cite{M2}. 

Surfaces which M-branes wrap in order to produce BPS states can be classified 
using the mathematical theory of calibrations. In \cite{Amherst} it was shown 
how the supergravity solution for a wrapped M-brane follows almost 
immediately once we are given the $p$-form which calibrates the wrapped cycle. 
This procedure is both simple and elegant but since it could only be used 
if the calibrations were already known, its applications were limited in
\cite{Amherst} to M-branes wrapped on Kahler calibrated surfaces 
in Kahler manifolds.

Using a different approach, the supergravity solution for an
M5-brane on a Riemann surface in a three complex dimensional manifold was 
found in \cite{FS2}. It turned out that the complex manifold need not be 
Kahler and is in fact defined through a somewhat unusual 
constraint on the Hermitean metric. By investigating its implications 
for calibrations in this background, we
are able to use the metric constraint to categorize a class of
non-Kahler calibrations which exist in spacetimes with non-zero flux. 

A complete classification of calibrations in such spacetimes will take substantial 
time and effort. We make a beginning here by studying the calibrations 
relevant to M-branes wrapping holomorphic cycles, as these are the 
simplest possible supersymmetric cycles one can consider. By exploiting the 
metric constraint referred to above, we are able to determine when non-Kahler 
calibrations arise and the conditions they are subject to. 

We start in Section \ref{BPS} with a lightning review of BPS brane configurations
in M-Theory. In Section \ref{calib} we introduce calibrations in purely geometric 
backgrounds first before explaining the modifications necessary in order to extend the
concept to include spacetime flux. In Section \ref{fssptime} we 
outline first the method employed by Fayyazuddin and Smith \cite{FS1} to find supergravity 
solutions for wrapped branes and then go on to show how the same results can 
be obtained more simply, a la \cite{Amherst}, using calibrations.

In Section \ref{genrule} we present the main result of the paper: a general rule 
which enables us to determine the constraint on the metric (or calibration)
corresponding to an M-brane wrapping a holomorphic cycle. This rule applies 
to both membranes and fivebranes, but since wrapped M2-branes have already
been discussed in depth in \cite{M2}, we concentrate mainly on M5-brane 
examples in section \ref{examples}. 
Section \ref{oddone} concludes with the reminder that though we have moved 
beyond Kahlerity and catagorized a new type of calibration, many such classes 
still live in un-named oblivion, waiting to be found!

\section{BPS Branes in M-Theory}
\label{BPS}

\subsection{Flat Branes}

We start by reviewing a few basic facts about M-branes.
In the expressions which follow, as indeed in the rest of the paper,
$X^{\mu}$ denotes coordinates tangent to a brane, $X^{\alpha}$ is
used to denote transverse coordinates and $r = \sqrt{X^{\alpha} X_{\alpha}} $
is the radial coordinate in this transverse space.\\

\no
\underline{\bf \sf The M5-brane:} A flat M5-brane with worldvolume
$X^{{\mu}_0} \dots X^{{\mu}_5}$ is
a half-BPS object which preserves 16 real supersymmetries corresponding
to the components of a spinor $\chi$ which satisfies the condition:
\be
{\hat \Gamma}_{{\mu}_0{\mu}_1{\mu}_2{\mu}_3{\mu}_4{\mu}_5} \chi = \chi.
\ee
In the presence of this brane, the geometry of spacetime is described by
the metric:
\bea
ds^2 &=& H^{-1/3} \eta_{\mu \nu} dX^{\mu} dX^{\nu} +
H^{2/3} \delta_{\alpha \beta} dX^{\alpha} dX^{\beta} \;\;\;\;\;\;\;\;\;\;\;\;\\
{\rm where}\;\;\;\;\;\;\;\;\;\;\;\;\;\;\;\;\;\;\; H &=& 1 + \frac{a}{r^4}.
\label{flatm5metric}
\eea
Since the M5-brane is charged under the supergravity three-form, it gives rise to a
four-form field strength:
\be
F_{\alpha \beta \gamma \delta} =
\frac{1}{2} \epsilon_{\alpha \beta \gamma \delta \rho}
\partial_{\rho} H.
\label{flatm54form}
\ee
Together, the equations (\ref{flatm5metric}) and (\ref{flatm54form}) specify the bosonic
fields in the supergravity solution for the M5-brane.\\

\no
\underline{\bf \sf The M2-brane:} A flat M2-brane spanning directions $X^{{\mu}_0}X^{{\mu}_1}
X^{{\mu}_2}$ is also a half-BPS object. Preserved supersymmetries
correspond to the 16 components of a spinor $\chi$ which survive the following projection
\be
{\hat \Gamma}_{{\mu}_0{\mu}_1{\mu}_2} \chi = \chi.
\ee
The bosonic content of the M2-brane supergravity solution is specified by the following expressions:
\bea
ds^2 &=& H^{-2/3} \eta_{\mu \nu} dX^{\mu} dX^{\nu} +
H^{1/3} \delta_{\alpha \beta} dX^{\alpha} dX^{\beta} \;\;\;\;\;\;\;\;\;\;\;\;\\
F_{{\mu}_0{\mu}_1{\mu}_2 \alpha} &=&
\frac{ \partial_{\alpha} H}{2 H^2}, \;\;\;\;\;\;\;\;\;\;\;\;\\
{\rm where}\;\;\;\;\;\;\;\;\;\;\;\;\;\;\;\;\;\;\; H &=& 1 + \frac{a}{r^6}.
\eea

\subsection{Intersecting Branes.}

One way of generating BPS states from the flat M-branes described above 
is to construct configurations of intersecting branes. In order for two M-branes 
to have a dynamic intersection, there must exist a worldvolume field to which 
the intersection can couple, either electrically or magnetically. 

The bosonic scalars in the three-dimensional theory on the membrane
worldvolume are dual to one-forms which can couple to point particles. Hence, a pair
of membranes should overlap only at a point. Assuming that the self dual three-form on 
an M5 brane vanishes, the only bosonic fields turned on in the worldvolume theory 
of the fivebrane are scalars. Two M5-branes must intersect with each other 
in 3 spatial dimensions, in order for their intersection to couple (magnetically) 
to a scalar and thus have worldvolume dynamics. These arguments lead to what is known as the 
(p-2) self intersection rule \cite{p-2} which states that
BPS states can be built from M-branes if these are oriented such that each pair 
of M$p$-branes has a $(p-2)$-dimensional spatial intersection . 

The Killing spinors of the resulting intersecting brane system must simultaneously 
satisfy the projection 
conditions imposed by each of its flat M-brane constituents. 
In general, a complex structure can be defined on the relative transverse
directions, i.e those which are spanned by some but not all of the constituent 
branes. The intersecting brane configuration can 
often be recovered as the singular limit of a single M-brane wrapping a 
smooth holomorphic cycle in this complex space.

\subsection{Wrapped Branes}

Another way to generate BPS states is to wrap an M-brane on a supersymmetric
$n$-cycle. It is known that holomorphic cycles are supersymmetric, and for the
time being, we will restrict ourselves to considering these alone. 

The amount of supersymmetry preserved by a particular spacetime can be found 
by introducing a brane which only probes the geometry and does not cause it to
deformation in any way; Killing spinors of the probe brane are then Killing 
spinors of the background, as long as the probe is placed parallel to the wrapped brane
and hence does not break any further supersymmetry. The Killing spinors of the probe brane 
satisfy the following projection condition \cite{BBS}:
\be
\chi = \frac{1}{p!} \epsilon^{\alpha_0\alpha_1 ... \alpha_p}
\Gamma_{M_0M_1 ... M_p} \partial_{\alpha_0} X^{M_0}
\partial_{\alpha_1} X^{M_1} ....
\partial_{\alpha_p} X^{M_p} \chi
\label{KSeqn}
\ee
where $\alpha_i$ are worldvolume indices, and $X^{M_i}$ describe the embedding of 
the $p$-brane in the ambient space-time. 
By virtue of being a spinor in 11 dimensions, $\chi$ must also be Majorana. This
requirement, when imposed alongside (\ref{KSeqn}) determines the 
number of supercharges preserved by the wrapped M-brane. 

It is worth bearing in mind that intersecting brane configurations are 
a subclass of the wrapped branes considered in this paper and are obtained in 
the limit when some cycles the brane wraps become singular. 

\section{Calibrations and Generalisations Thereof.}
\label{calib}

Calibrations $\phi_p$ are a very useful mathematical construction
which enable us to classify minimal $p$-dimensional submanifolds
in a given spacetime \cite{HarveyLawson}. A closed $p$-form $\phi_p$ is called a
{\bf (standard) calibration} if it obeys the inequality
\be
|{\phi_p}_{|_{{\Lambda}_p}}| \; \leq \; dV_{{\Lambda}_p}
\label{caldefn}
\ee
for all $p$-dimensional submanifolds ${\Lambda}_p$. 
A manifold $\Sigma_p$ which saturates the inequality is
known as a minimal or calibrated manifold. 

In the absence of space-time flux, branes ensure their stability by minimizing their
worldvolumes; the volume form of an uncharged stable brane
must therefore be a calibration in spacetime\footnote{While the
expressions we have used to illustrate these statements are taken from
11-dimensions, the same logic and results apply to lower dimensions as 
well.}. In such purely geometric backgrounds, Killing spinors are covariantly 
constant. 

If the brane is charged, we can take our pick of two alternate ways to 
describe the ambient spacetime. The bosonic fields needed to specify
the background are the metric and the flux of the gauge field which
couples to the brane. The first and most obvious attitude we can take 
is to treat these fields as two seperate objects.    
Killing spinors of the brane configuration are determined by the 
metric as well as the field strength and are hence no longer covariantly constant. 
Since the coupling of the gauge potential to the brane must now be taken 
into account, it is only natural that the criterion for brane
stability also should change. In fact it turns out that stability
requires now that the energy of the brane be minimized, where the energy is
a measure not only of the volume but also the charge. 

Alternately, we could adopt a more unconventional point of view and
'combine' the effects of the metric and the flux into a suitably
redefined metric. This new metric is defined in such a way that it
sees Killing spinors as being covariantly constant. By shifting the effect of
the field strength into the geometry, we have of course made the
geometry substantially more complicated and the new metric will in
general have torsion. 

To reiterate then, the volume-form of a stable charged brane is
minimal only when measured by a metric which has been defined so as to
incorporate the effect of the flux.
{\bf Generalised calibrations} $\phi_p$ are thus defined such that
\bea
d (A_{p} + \phi_p) &=& 0 \\
{\rm and} \;\;\;\;\;\;\; | {\phi_p}_{|\Sigma_p} |
\; &\leq& \; \tilde{dV}_{\Sigma_p}
\label{gencaldefn}
\eea
for any $p$-dimensional submanifold ${\Sigma_p}$, where $\tilde{dV}$
is the volume form in terms of the redefined metric with torsion.
It is also clear from the above that a generalised calibration is not closed,
but rather is gauge equivalent to the potential $A_p$ under which the
$(p-1)$brane is charged.

Typically, only those directions along the brane worldvolume which are wrapped 
on a supersymmetric curve have a non-trivial space-time embedding; the 
remaining directions are flat\footnote{In 
the language of this paper, these are the $X^{\mu}$}. Had its entire world-volume
been flat, the brane would be a 1/2 BPS object; wrapped branes however, generically
break more than half the supersymmetry. Since the preserved
supercharges depend on the geometry of the supersymmetric cycle, it is
the wrapped directions of the brane world-volume which play a
essential role in this analysis whereas flat directions contribute trivially. 
Because of this, 
the volume form of a BPS brane, which is actually a calibration in the full 
spacetime, is sometimes referred to as a calibration even in the subspace 
where the cycle is embedded. 

\bigskip
\begin{center}
\noindent\fbox {\noindent\parbox{5.5in}{ {\sf \bf Clarifying Our Conventions:}\\

The definition of a generalised calibration in the embedding space ${\cal M}$
adopted here differs from that in \cite{GenCal}. In order to avoid any
confusion in a comparison of the two papers, let us make this difference explicit. 

A generalised calibration $\Phi_{p+1}$in 11 dimensions is gauge equivalent to the
potential $A_{p+1}$ to which the M$p$-brane couples electrically. We can split this up
into the product of an $l$-form $\lambda_{l}$ (where $l = p+1 -m$) along the 
flat 
directions and an $m$-form $\phi_m$ in ${\cal M}$, such that both are 
generalised calibrations in their own subspaces. The components of these forms 
satisfy the relation:
\be
\Phi_{i_0 \dots i_{p+1}} = \lambda_{i_0 \dots i_{l}} \times \phi_{i_{l+1} \dots i_{p+1}}
\equiv A_{i_0 \dots i_{p+1}}
\ee

In \cite{GenCal}, it is simply the gauge potential restricted to ${\cal M}$ which 
is called the generalised calibration $\tilde{A}_m$ in this subspace. 
While this is of course still an $m$-form, its components differ from those of $\phi_m$,
and can in fact be obtained from $A_{i_0 \dots i_{p+1}}$ merely by holding the first
$l$ indices fixed, as follows:
\be
\tilde{A}_{i_1 \dots i_m} = A_{1 2...l,i_1 \dots i_m} 
\ee}}
\end{center}
\bigskip

\section{Supergravity Solutions for Wrapped Branes.}
\label{fssptime}

Supergravity solutions for intersecting brane configurations were constructed
initially using the harmonic function rule \cite{Tseytlin}, which is a recipe 
for combining the supergravity solutions of each component M-brane; the resulting 
solutions however were smeared along the relative transverse space. 
In \cite{FS1}, Fayyazuddin and Smith came up with a metric ansatz which 
enabled them to consider localised brane intersections. 
Since the method employed there will be used later in this paper, here is a brief overview of 
the way it works. 

\subsection{Fayyazuddin-Smith Spacetimes}
\label{fsmethod}

We begin by postulating a form for the spacetime metric, based on 
isometries of the wrapped brane configuration. Since we are 
focusing on bosonic backgrounds with Killing spinors, the
gravitino has already been set to zero and supersymmetry preservation can be ensured 
simply by demanding that the gravitino variation vanish as well. This is imposed by hand 
and leads
to a set of constraints on the field strength of the supergravity three form, as well
as on the functions in the metric ansatz. If in addition, the 
field strength obeys the Bianchi Identity and equations of motion, it is 
guaranteed that Einstein's equations will be satisfied. Having outlined the method, 
we discuss a few of its steps in more detail in order to 
equip ourselves to apply the procedure to explicit examples.\\ 

\no
\underline{\sf Metric Ansatz:} 
We want to consider a class of spacetimes which describe M-branes wrapping
supersymmetric cycles. Depending on the dimension of the cycle, there could be some 
world-volume directions $X_{\mu}$ which are left unwrapped or flat; Lorentz invariance should 
be preserved along these directions. If the supersymmetric 
cycle in question is embedded into an $m$ dimensional subspace spanned by $X_a$, then 
the remaining directions of spacetime $X_{\alpha}$, (those transverse to both 
$X_{\mu}$ and the embedding space), should be rotationally invariant.  
A metric describing this 11-dimensional spacetime takes the form:
\be
ds^2  = {H_1}^2 \eta_{\mu \nu} dX^{\mu} dX^{\nu} + h_{a b} dX^a dX^b
+ {H_2}^2 \delta_{\alpha \beta} dX^{\alpha} dX^{\beta}
\ee
From our discussion about the isometries of spacetime, we can see that the 
metric must be diagonal in both $X_{\mu}$ and $X_{\alpha}$. However we cannot 
say anything about the metric in the embedding space yet beyond a comment that 
it must, along with $H_1$ and $H_2$, be independent of $X_{\mu}$. 

Since we are going to be concerned only with M-branes wrapping holomorphic cycles,
it is convenient to define a complex structure on the embedding space. This makes the 
holomorphicity of the cycle (and therefore the supersymmetry of the configuration) manifest. 
We can hence replace the embedding space metric $h_{a b} dX^a dX^b$ 
with $2 G_{M {\bar N}} dz^{M} dz^{\bar N}$, arriving at the following final form for the metric
\be
ds^2  = {H_1}^2 \eta_{\mu \nu} dX^{\mu} dX^{\nu} +
2 G_{M {\bar N}} dz^{M} dz^{\bar N} + {H_2}^2
\delta_{\alpha \beta} dX^{\alpha} dX^{\beta}
\label{standard}
\ee
which will be used throughout this paper.\\

\no
\underline {\bf \sf Killing Spinors:} 
A Majorana spinor in the 11-dimensional background described above
can be decomposed into components along and transverse to the complex subspace where the 
supersymmetric cycle is embedded. The component which lies in the complex 
space can be expressed by a linear combination of Fock space states using
the fact that the Clifford algebra in $\C^n$ takes on the form of an 
algebra of $n$ fermionic creation and annihilation operators. The creation operators 
can then act on the vacuum to generate a Fock space in which states are labelled by 
$n$ fermionic occupation numbers; one corresponding to each creation operator.\\ 

\no
\underline {\bf \sf Supersymmetric Variation of the Gravitino:} 
The Killing spinor for a particular brane configuration can be decomposed as 
described above and substituted in the gravitino variation equation:
\be
\delta_{\chi}\Psi_{I} = (\partial_{I}  + \frac{1}{4} \omega_I^{ij} \hat{\Gamma}_{ij}
+ \frac{1}{144}{\Gamma_{I}}^{JKLM}F_{JKLM}
-\frac{1}{18}\Gamma^{JKL}F_{IJKL})\chi.
\label{susy}
\ee
The requirement $\delta_{\chi} \Psi = 0$ leads to the vanishing of a 
combination of Fock space states. Since these states are linearly 
independent, the coefficient of each must be set to
zero identically. This gives us a set of constraints on the metric and field strength.

\subsection{A Shortcut ..}
\label{shortcut}

From our earlier discussion of calibrations, we have learnt that 
the generalised calibration $\phi$ corresponding to a wrapped brane is gauge equivalent 
to the $(p+1)$-form potential to which the brane couples. 
The field strength $F= dA = d\phi$, follows once we are 
given a suitable calibration. For the spacetimes described in the previous section, 
the calibrating
form, and hence the field strength, can simply be read off from the metric! The entire
bosonic content of the supergravity solution can thus be obtained simply by appealing
to the isometries of the configuration and a knowledge of the possible calibrations.    

In order to make this construction more explicit\footnote{
Details are given in \cite{GenCal} and \cite{Amherst}.}, we apply it to the example of
a membrane wrapping a 2-cycle embedded holomorphically into $\C^3$.\\

\no
\underline {\bf \sf M2 on a holomorphic curve in $\C^3$}

\no
The Fayyazuddin-Smith ansatz for the metric is:
\be
ds^2 = - H^{2/3}  dt^{2} + 2 G_{M {\bar N}} dz^{M}
dz^{\bar N} + H^{1/3} \delta_{\alpha \beta} dX^{\alpha} dX^{\beta}.
\label{eq:standard}
\ee
where $z^{M}$ are complex coordinates, $X^{\alpha}$ span the 
remaining 4 transverse directions and supersymmetry requirements have been used 
to fix two of the undetermined quantities in the metric ansatz in terms of a single
harmonic function H. Given this metric, we can immediately read off 
the calibrating three-form 
\be
\Phi_{0 M \bar{N}} = i \sqrt{H^{2/3}} dt \wedge G_{M {\bar N}} dz^{M}
\wedge dz^{\bar N}
\ee
and use it to compute the components of the field strength:
\be
F_{I 0 M \bar{N}} = d_{[I,} \Phi_{0 M \bar{N}]} 
\ee
where I can be any spatial coordinate. 

In \cite{Amherst}, where this method was proposed, it was assumed 
(for lack of any information to the contrary) that the metric 
$G_{M {\bar N}}$ is Kahler. However, as we shall soon see, this is by no means 
the only possibility! 

\subsection{... and its Limitations!}

Though calibration technology provides us with a procedure both quicker and
simpler than the one described in section \ref{fsmethod}, the fact remains that 
in order to write down supergravity solutions for M-branes wrapped on calibrated 
manifolds, we need to know first what the calibrated forms are! Uptil now, this 
method had only been used for M-branes wrapped around Kahler calibrations
\cite{Amherst} simply because these were the only calibrations we were aware 
of in spacetimes with non-trivial four-form flux, so the full potential 
which lay latent in this method could not be exploited.

This is a problem of larger proportions than one might at first think. While it 
is true that Kahler manifolds permeate String and M-Theory, it is also a fact that
we frequently encounter non-Kahler manifolds, perhaps without recognizing them 
as such. In order to susbtantiate this statement, we take a short detour 
through a brane described by a non-Kahler calibration, before
proceeding to discuss the general constraint which can be used to identify 
a class of calibrations in M-Theory backgrounds.\\      

\no
\underline{\sf \bf Non-Kahler Calibrations.}:

\no
A flat M5-brane can be thought of as being wrapped on a trivial supersymmetric
cycle embedded into a subspace of 11 dimensional spacetime. Take this 
subspace to be $\C^3$, spanned by holomorphic coordinates 
$u, v, w$. Two equations $f = f(u,v,w) = 0$ and $g = g(u,v,w) = 0$ are then needed 
to define a holomorphic two-cycle. If these equations are $v = 0$ 
and $w = 0$, the two-cycle in question is simply the complex $u$ plane. 

In the presence of a flat M5-brane with worldvolume $0123u\bar{u}$, the 
spacetime metric is given by: 
$$ds^2 = H^{-1/3} (- dt^2  +  dx_1^2 + dx_2^2 + dx_3^2 + du d{\bar u}) + 
H^{2/3} (dv d{\bar v} + dw d{\bar w} + dy^2) $$
where 
$$ H = \frac{{\rm constant}}{(|v|^2 + |w|^2 + y^2)^{3/2}}$$

In general, an M5-brane wrapping a Riemann surface embedded in $\C^3$ is expected
to give rise to a metric of the form \cite{FS2}:
$$ds^2 = H^{-1/3} (- dt^2  +  dx_1^2 + dx_2^2 + dx_3^2) + 2 G_{M {\bar N}} 
dz^M dz^{\bar N} + H^{2/3} dy^2 $$
where $0123$ are the flat directions of the brane and $y$ is the single coordinate
transverse to both the brane and the complex space. Comparing the two 
expressions, we can read off the components of the Hermitean
metric: 
$$2 G_{u {\bar u}} = H^{- 1/3} \; \; , \; \; 2 G_{v {\bar v}} =  H^{2/3}
\; \; , \; \; 2 G_{w {\bar w}} = H^{2/3}.$$ 
It is convenient to define a rescaled metric, 
$g_{M {\bar N}} = H^{-1/6} G_{M {\bar N}}$ whose components are
$$2 g_{u {\bar u}} = H^{- 1/2} \; \; , \; \; 2 g_{v {\bar v}} =  H^{1/2}
\; \; , \; \; 2 g_{w {\bar w}} = H^{1/2}.$$ 
It can trivially be seen that this metric is not Kahler; moreover
since a Kahler metric cannot be obtained even by rescaling, 
$g_{M {\bar N}}$ is not warped Kahler either. However, the components 
of this blatantly non-Kahler metric satisfy the following curious relations: 
\be
\partial_u g_{v {\bar v}} g_{w {\bar w}} =
\partial_v g_{u {\bar u}} g_{w {\bar w}} =
\partial_w g_{u {\bar u}} g_{v {\bar v}} = 0.
\ee
In terms of the Hermitean form 
$\omega = i g_{M {\bar N}} dz^M dz^{\bar N}$ associated with the metric, 
this can be re-expressed as follows:
$$\partial [\omega \wedge \omega] = 0 \; \; {\rm but} \; \; \partial \omega 
\neq 0 $$ 

As we will see in the following sections, this M5-brane is by no means the 
sole example of commonly encountered non-Kahler calibrations!

\section{A Method to the Madness.}
\label{genrule}

In our search for non-Kahler calibrations, we don't exactly have to look very far.
Apart from the M5 example quoted above, there are membranes aplenty willing to oblige!

\subsection{The Madness}

Supergravity solutions for a class of BPS states corresponding to 
wrapped membranes were dicussed in \cite{M2}. In keeping with the logic 
that holomorphicity implies supersymmetry, the M2-branes were wrapped on 
holomorphic cycles in complex subspaces of varying dimension $n$.
Working with the following (standard) ansatz for the spacetime metric:
\be
ds^2 = - H^{-2/3}  dt^{2} + 2 G_{M {\bar N}} dz^{M}
dz^{\bar N} + H^{1/3} \delta_{\alpha \beta} dX^{\alpha} dX^{\beta},
\label{eq:standard}
\ee
where $z^{M}$ are $n$ holomorphic coordinates and 
$X^{\alpha}$ span the $(10 - 2n)$ transverse directions, 
it was found that in each case, supersymmetry preservation imposes a constraint 
on the Hermitean metric. 
These constraints take the following form:
\bea
\partial [H^{2/3} \omega] &=& 0 \;\;\; {\rm for} \; n=2 \nonumber\\
\partial [H^{1/3} \omega \wedge \omega] &=& 0 \;\;\; {\rm for} \; n=3\nonumber\\
\partial [\omega \wedge \omega \wedge \omega] &=& 0 \;\;\; {\rm for} \; n=4 \nonumber\\
\partial [H^{- 1/3} \omega \wedge \omega \wedge \omega \wedge \omega] &=& 0 \;\;\; {\rm for} \; n=5
\label{m2metrics}
\eea
where $\omega = i G_{M {\bar N}} dz^{M} \wedge dz^{\bar N}$ is the Hermitean two-form 
associated with the metric in the complex subspace. 

While (warped) Kahler metrics definitely solve the above constraints they by no means 
exhaust the available options. Hence, by assuming that the metric on the 
embedding space is Kahler, we are 
in fact restricting ourselves unnecesarily and losing out on a wealth of possibilities. 

We have now discussed two ways of obtaining supergravity solutions. Each has its merits. 
The first, explained in section \ref{fsmethod}, involves somewhat lengthy 
computations but includes in its results a statement about the kind of manifold 
the embedding space must be. The second procedure, outlined in section \ref{shortcut}, 
is quick and simple  
but it does not yield a restriction on the embedding space metric; this must be provided as 
an external input.  

Poised between the labour of one method, and the simplicity of the other, it is inevitable to 
ask the question, can we somehow combine the best features of both approaches? 
Is it possible to
broaden the class of calibrations under consideration so that we have the ease of writing down
supergravity solutions using calibrations and also the assurance that all possible cases 
are covered? 

\subsection{The Method.}

A persistent feature of the wrapped brane supergravity solutions obtained 
using the method of \cite{FS1} is a constraint on the metric in the subspace 
where the supersymmetric cycle is embedded. This constraint in turn restricts the 
$(p+1)$-form potential to which the brane couples. Since the 
potential is gauge equivalent to the generalised calibration, we find that 
in fact the metric constraint can alternately be viewed as a condition which determines 
generalised calibrations in the given background. 

We propose the following:\\

{\sf
\no
In 11-dimensional backgrounds with non-zero four-form flux, 
a class of generalised calibrations in the embedding space {\cal M} is given by the 
$2m$-forms $\phi_{2m}$, for
\bea
\begin{array}{ccc}
\phi_{2m} & = & (\underbrace{\omega \wedge \omega \wedge ..... \omega \wedge \omega})\\
 &  & m
\end{array}
\eea
if the following constraint holds:
\be
\partial *_{\cal M} [ \phi_{2m} |G^{'}|^{1/2m} ] = 0.
\label{fcond}
\ee
Here, $G^{'}$ denotes the determinant of the metric restricted to directions transverse 
to the embedding space, and the Hodge dual is taken within the embedding space.} \\

\section{Proof by Example.}
\label{examples}

The aim of this section is to illustrate the exhaustive nature of the statement 
(\ref{fcond}) through a
case by case analysis of M-branes wrapping holomorphic cycles in complex manifolds of
various dimensions.
 
\subsection{M2-branes on 2-cycles}

Two cycles are the only holomorphic curves on which M2-branes can be wrapped.
In an 11-dimensional spacetime, these curves can be embedded in manifolds of 
complex dimensions $2 3, 4$ and $5$. 
The supergravity solutions corresponding to each of these configurations were studied in 
\cite{M2} and the metric constraints that accompany each solution 
were reviewed in (\ref{m2metrics}). Notice now that they can be unified into 
the following expression:
\be
\partial *_{\C} [ \omega_{M {\bar N}} \sqrt{{\rm det} G^{'}} ] = 0
\label{m2rule}
\ee 
where ${\rm det} G^{'}$ is the determinant of the metric when restricted to directions
transverse to the complex sub-manifold. 

The part of the gauge potential to which the membrane worldspace couples is
given simply by $\omega_{M {\bar N}}$. In order for the membrane to be stable and 
supersymmetric, this two form should be a generalised calibration in the complex 
space. Hence, we find that the constraint (\ref{m2rule}) is but a special case of the 
general rule (\ref{fcond}) as applied to M2-branes.\\ 

\subsection{M5-branes on 2-cycles}

In this section we employ the calibration approach to find supergravity solutions
for fivebranes wrapped on holomorphic two-cycles, using calibrations determined 
by the constraint (\ref{fcond}). This procedure enables us to 
reproduce in a few simple steps the results of \cite{FS1} and \cite{FS2}.\\

\newpage

\no
\underline{\sf \bf M5 on a holomorphic curve in $\C^2$}:

When an M5-brane wraps a holomorphic submanifold in $\C^2$,
the relevant ansatz for the spacetime metric is:
\be
ds^2 = H^{- 1/3}  \eta_{\mu \nu} dX^{\mu} dX^{\nu}
+ 2 G_{M {\bar N}} dz^{M} dz^{\bar N} +
H^{2/3} \delta_{\alpha \beta} dX^{\alpha} dX^{\beta}.
\ee
where $z^{M}$ are coordinates on $\C^2$, $\alpha$ takes values $8,9,10$
and $\mu$ runs over $0,1,2,3$.
Supersymmetry preservation has already been used to fix the relative coefficients 
$H^{- 1/3}$ and $H^{2/3}$. It further dictates that, (upto rescaling by an
arbitrary holomorphic function), the harmonic function H is
related to the determinant G of the Hermitan metric by $${\sqrt G} = H^{2/3}.$$
The calibrating form of the M5-brane
\bea
\Phi &=& H^{-2/3} G_{M {\bar N}}
dt \wedge dX^1 \wedge dX^2 \wedge dX^3 \wedge
dz \wedge dz^{\bar{N}}\\
&=& dV_{0123} \wedge \phi_{M \bar{N} }
\label{vfm22}
\eea
must be such that:
\be
\partial [H^{1/3} G_{M {\bar N}}] = 0.
\ee
If the above condition is satisfied, the generalised calibration
is gauge equivalent to the potential $A$ under which the brane is charged.
The only non-vanishing component of this potential can then be read off
from (\ref{vfm22}) as:
\be
A_{0123 M \bar{N} } = H^{-2/3} G_{M \bar {N}}
\ee
Consequently, the supergravity four form $F_4 = * d A$ is given by:
\bea
F_{M 8 9 (10)} &=& - \frac{i}{2} \partial_{M} H \\
F_{\bar{N} 8 9 (10) } &=& \frac{i}{2} \partial_{\bar{N}} H \\
F_{N \bar{M} \beta \gamma} &=& \frac{i}{2} \epsilon_{\alpha \beta \gamma}
\partial_{\alpha} [H^{1/3} G_{N \bar{M}}]
\eea
These results agree exactly with those obtained in \cite{FS1}.\\

\no
\underline{\sf \bf M5 on a holomorphic curve in $\C^3$}:

When the M5brane is wrapped on a holomorphic curve embedded in $\C^3$,
the metric takes the form:
\be
ds^2 = H^{- 1/3}  \eta_{\mu \nu} dX^{\mu} dX^{\nu}
+ G_{M {\bar N}} dz^{M} dz^{\bar N} +
H^{2/3} dy^2.
\ee
where $z^{M}$ now span $\C^3$, $y$ is the single overall transverse direction.
and the harmonic function H is related to the determinant of the Hermitean
metric by $H = \sqrt{G}$.
In this background, the wrapped M5-brane is calibrated by the form
\bea
\Phi &=& H^{-2/3} G_{M {\bar N}} dt \wedge dX^1 \wedge dX^2 \wedge dX^3 \wedge
dz \wedge dz^{\bar{N}}\\
&=& dV_{0123} \wedge \phi_{M \bar{N} }
\label{vfm23}
\eea
such that
\be
\partial [H^{- 1/3} \omega_G \wedge \omega_G] = 0.
\ee
When this relation holds, the gauge potential $A$ is once more given by
\be
A_{0123 M \bar{N} }  = H^{-2/3} G_{M \bar {N}}
\ee
This leads to the following expressions for the four-form field strength:
\bea
F_{\bar{N} \bar{P} M y} &=& \frac{1}{2}
[\partial_{\bar{P}} (H^{1/3} G_{M \bar{N}}) -
\partial_{\bar{N}} (H^{1/3} G_{M \bar{P}})], \\
F_{N P \bar{M} y} &=& \frac{1}{2}
[\partial_{P} (H^{1/3} G_{N \bar{M}}) -
\partial_{N} (H^{1/3} G_{P \bar{M}})], \\
F_{M N \bar{P} \bar{Q}} &=& \frac{i}{2} \partial_{y}
[H^{-1/3} (G_{M \bar{Q}} G_{N \bar{P}} - G_{M \bar{P}} G_{N \bar{Q}})]
\eea
which are precisely the expressions obtained in \cite{FS2}. 

\subsection{M5-branes on 4-cycles}

The worldvolumes of fivebranes are large enough to be wrapped on holomorphic 
four-cycles as well. The smallest complex space into which a four-manifold can have a
non-trivial holomorphic embedding is $\C^3$, and this is the configuration we turn to now.
As we will see later, this is in fact the only M-brane which can be wrapped on a 
holomorphic four-cycle within the scope of this paper.\\ 

\no
\underline{\sf \bf The Fayyazuddin-Smith Treatment}: \\

\no
\underline{\sf Metric Ansatz:}
\no
For an M5-brane wrapped on a 4-cycle $\Sigma_4$
in $\C^3$, the metric takes the form:
\be
ds^2 = H_1^{2}  \eta_{\mu \nu} dX^{\mu} dX^{\nu}
+ 2 G_{M {\bar N}} dz^{M} dz^{\bar N} +
H_2^{2} \delta_{\alpha \beta} dX^{\alpha} dX^{\beta}
\ee
where $\mu=0,1$ labels the unwrapped directions, $z^{M}$ are
holomorphic coordinates in $\C^3$ and $\alpha$ takes values 8,9, and 10.\\

\no
\underline{\sf Killing Spinors:}
\no
The Killing spinors in this 
space-time are given by (\ref{KSeqn}): 
\be
\epsilon^{abcd}
\Gamma_{0 1} \Gamma_{m \bar{n}  p \bar{q}}
\partial_{a} X^{m} \partial_{b} X^{\bar{n}}
\partial_{c} X^{p} \partial_{d} X^{\bar{q}} \chi =
\chi
\ee
where the $\Gamma_m$ are flat space $\Gamma$-matrices and ${\sigma}^a \dots {\sigma}^d$ 
are coordinates on the four-cycle. This leads to the condition
\be
\Gamma_{01}\Gamma_{ m \bar{n} p \bar{q}} \chi =
(\eta_{m \bar{n}} \eta_{ p \bar{q}} -
\eta_{m \bar{q}} \eta_{p \bar{n}} )
\chi
\ee
where $\eta_{m \bar{n}}$ is the flat space metric. A solution is given by
\be
\chi = a\otimes|000> + b\otimes|111>
\label{chi}
\ee
if the spinors $a$ and $b$ in the $(1+4)$ dimensional space-time
transverse to $\C^3$ satisfy:
\bea
\Gamma_{89(10)} \; a &=& -i a \nonumber \\
\Gamma_{89(10)} \; b &=& \; i b
\label{KS2}
\eea
The wrapped M5-brane then preserves $\frac{1}{8}$ of the spacetime
supersymmetry, corresponding to the 4 spinors which satisfy the
above conditions.\\

\no
\underline{\sf Supersymmetric Variation of the Gravitino:} 
\no
The BPS supergravity solutions for the wrapped brane are found by 
demanding that $\delta_{\chi} \Psi = 0$ holds 
for a metric of the form (\ref{standard}), when the variation parameter
$\chi$ is given by (\ref{chi}). This gives rise to the following set of equations:
\be
3 \partial_{\bar{M}} ln \; H_1 - i  H_2^{-3} F_{{\bar{M}} 8 9 (10)} = 0
\ee
\be
3 \hat{\epsilon}_{\alpha \beta \gamma} (\partial_{\gamma} H_2 )
+ 2i G^{M {\bar{N}}} F_{M {\bar{N}} \beta \gamma} = 0
\ee
\be
3 G^{N {\bar{P}}} ( \partial_{\bar{P}} G_{N {\bar{M}}} ) +
4iH_2^{-3} F_{{\bar{M}} 8 9 (10)} = 0
\ee
\be
6 i H_2 \partial_{\alpha} ln \; H_1 + \hat{\epsilon}_{\alpha \beta \gamma}
G^{M {\bar{N}} } F_{M {\bar{N}} \beta \gamma} = 0
\ee
\be
i H_2^{-1} [6  F_{M {\bar{N}} \beta \gamma} -
2 G_{M {\bar{N}}} G^{P {\bar{Q}}} F_{P {\bar{Q}} \beta \gamma}] + 3
\hat{\epsilon}_{\alpha \beta \gamma}
(\partial_{\alpha} G_{M {\bar {N}}} ) = 0
\label{detzero}
\ee
\be
3H_2^{3} (\partial_{\bar {P}} G_{M {\bar{N}}} - \partial_{\bar{N}}
G_{M {\bar{P}}} )
- 2i G_{M {\bar{N}}} F_{{\bar{P}} 8 9 (10)} + 2i G_{M {\bar{P}}}
F_{{\bar{N}} 8 9 (10)} = 0
\ee

\bigskip
\begin{center}
\noindent\fbox {\noindent\parbox{5.5in}{{\sf \bf The Orginal Variables: An Aside}\\

Formulating the solutions in terms of $H_1, H_2$ and $G_{M {\bar{P}}}$,
we find that the harmonic functions, though related to each other as expected,  
are independent of the determinant, G. Moreover, 
$$\partial_{\alpha} G = \partial_{M} G = \partial_{\bar{N}} G = 0.$$
Though this might seem puzzling at first, it is important to realise that we
are not implying that the Hermitean metric is independent of the spatial coordinates; 
merely that such dependences cancel out in its determinant! That this is infact
to be expected can be seen by looking at a flat M5 brane, trivially
embedded in  $\C^3$.\\

{\bf \sf Example:} For an M5 spanning 01$u {\bar u} v {\bar v}$, 
the supergravity solution is given by
$$ds^2 = H^{-1/3} (- dt^2  +  dx_1^2 + du d{\bar u} + dv d{\bar v}) + 
H^{2/3} ( dw d{\bar w} + dx_7^2 + dx_8^2 + dx_9^2 + dx_{10}^2), $$
enabling us to read off the following: 
\bea
H_1^2 = H^{-1/3} \; , \; H_2^2 = H^{2/3} 
&\Rightarrow& \; \; H_1^2 = H_2^{-1}  \; {\rm and }\; \nonumber \\
2 G_{u {\bar u}} =  2 G_{v {\bar v}} = H^{-1/3}  \; , \; 
2 G_{w {\bar w}} = H^{2/3} \; &\Rightarrow& \; 
{\sqrt G} = G_{u {\bar u}} G_{v {\bar v}} G_{w {\bar w}} = 1/8
\nonumber
\eea
Since the determinant of the Hermitean metric is a constant, its derivatives obviously 
vanish.}}

\end{center}
\bigskip

\no
In terms of the rescaled quantities:
\be
H \equiv H_2^3 \;\;\;\;\; {\rm and} \;\;\;\;\;
g_{M {\bar{N}}} \equiv H_2 G_{M {\bar{N}}}
\ee
solutions to the above equations are:
\bea
\partial_{\bar{N}} \; ln H &=& -6 \partial_{\bar{N}} \; ln H_1
= 3 \partial_{\bar{N}} \; ln H_2\\
\partial_{\bar {P}} g_{M {\bar{N}}} &=& \partial_{\bar{N}} g_{M {\bar{P}}} \\
F_{8 9 (10) M} &=& \frac{i}{2} \partial_{M} H \\
F_{N \bar{M} \beta \gamma} &=& \frac{i}{2} \epsilon_{\alpha \beta \gamma}
\partial_{\alpha} g_{N \bar{M}}
\eea
Since the supergravity three-form couples magnetically to the M5-brane,
$d * F = 0$ trivially, whereas the Bianchi Identity constrains the metric
to obey the following non-linear differential equation
\be
\partial^2_{\alpha} g_{M \bar{N}}  + 2 \;
\partial_{M} \partial_{\bar{N}} H = 0
\ee
This is precisely the solution found in \cite{Amherst}. What we have gained through
the above analysis is the metric constraint which explicitly rules out the
possibility of M5-branes being wrapped on non-Kahler holomorphic four-cycles
in three complex dimensional manifolds and tells us that the previously calculated
solution \cite{Amherst} is, in this case, the only option!\\

\no
\underline{\bf \sf The Calibration Method:}

The only non-vanishing component of the gauge potential corresponding to
an M5-brane wrapping a holomorphic four-cycle in $\C^3$ is
\be
\tilde{A}_{01 M N {\bar P} {\bar Q} } = H^{-1/3}
(G_{M {\bar P}} G_{N {\bar Q}} - G_{M {\bar Q}} G_{N {\bar P}})
\ee
Decomposing this into a two-form $dV_2$ along the $(0,1)$ directions
and a four-form $\phi_{M N {\bar P} {\bar Q}}$ in $\C^3$, we find that
the gauge-potential/world-volume is a calibration in space-time
only if $\phi_{M N {\bar P} {\bar Q}}$ is a generalised calibration in the
embedding space, i.e. it satisfies (\ref{fcond}). For an M5-brane wrapped
on a holomorphic $4$-cycle in $\C^3$, this imples that
\be
\partial *_{\C} [H^{1/3} \omega_G \wedge \omega_G] =  \partial \omega_g  = 0
\ee
where $\omega_G$ and $\omega_g$ are the Hermitean two-forms associated
with the metrics $G_{M {\bar N}}$ and $g_{M {\bar N}}$ respectively.
The field strength of the supergravity three form can now be easily
calculated using $F_4 = * d A_6$, reproducing previous results.

\section{The Odd One Out {\it or} What Lies Ahead.}
\label{oddone}

In order for a wrapped M-brane to be a stable, supersymmetric configuration the
gauge potential to which it couples must 
be (equivalent to) a generalised calibration in 11-dimensional space-time. In this paper,
we looked at M-branes wrapping holomorphic curves and found that when the  
background contains a field strength, supersymmetric cycles/calibrated forms 
in a complex subspace have a non-trivial dependence on the remaining directions of 
spacetime, as reflected in the constraint (\ref{fcond}). 

The examples considered here, together with those in  
\cite{M2} exhaust all possible cases of M-branes wrapped holomorphic cycles 
in complex subspaces of 11-dimensional space-time; with one notable exception. 
An M5 wrapped on a four-cycle in $\C^4$ has not been discussed and in fact its 
supergravity solution cannot be constructed based on a calibration following from 
(\ref{fcond}). We justify its exclusion from the present analysis by 
pointing out how it differs from the configurations we have looked at so far.

The definition of generalised calibrations presented in section \ref{calib}, 
and as a result the entire formalism of this paper, is restricted by 
construction to apply to configurations with a non-vanishing spacetime flux 
and no non-trivial fields on the brane world-volumes \footnote {A suitable 
modification of the calibration concept does exist which incorporates 
worldvolume fluxes \cite{wvflux}, however it has not been undertaken here.}. 
The rule (\ref{fcond}) comes with the in-built assumption
that the only bosonic fields turned on in the M-brane worldvolume theory are scalars. As 
a result, it applies to {\it only} to those wrapped M-branes which, in the 
intersecting brane picture can be interpreted as a system of 
branes where each pair has a $(p-2)$ dimensional spatial overlap.
 
It should by now be clear why (\ref{fcond}) cannot be applied to 
an M5 wrapping a four-cycle in $\C^4$. In the intersecting brane picture obtained
by taking the limit where the four-cycle becomes singular, the two M5-branes have only 
one common spatial dimension, thus violating the $(p-2)$ self intersection rule!

Though, by considering a particular type of BPS wrapped M-branes, we have succeeded 
in shedding light on a new catagory of calibrations, many still lurk in the shadows. 
This message is reinforced by the example discussed above. It is hoped that a study 
of this brane configuration will bring another class of calibrations into the spotlight.  
This theme will be expanded upon further in a paper coming soon to an arXiv near you! 

\acknowledgments{
Ansar Fayyazuddin has my gratitude for always being on hand with advice,
support and answers. I would like to acknowledge useful conversations
with Jerome Gauntlett, Chris Hull and Stathis Pakis during the Clay
School at the Newton Institute last year. I am grateful to the Theory
Group at the University of Texas at Austin for their hospitality and
to Amer Iqbal and Julie Blum for discussions about this and related work.}

\end{document}